# Weather Radar in Nepal: Opportunities and Challenges in Mountainous Region

Rocky Talchabhadel[1], Ganesh R. Ghimire[2], Sanjib Sharma[3], Piyush Dahal[4], Jeeban Panthi[5], Rupesh Baniya[6], Jayaram Pudashine[7], Bhesh Raj Thapa[8], Shakti PC[9], Binod Parajuli[10]*

[1]Disaster Prevention Research Institute, Kyoto University, Fushimi-ku, Kyoto, 612-8235, Japan; rocky.ioe@gmail.com
[2]IIHR Hydroscience and Engineering, The University of Iowa, Iowa City, Iowa, USA; ganeshghimire1986@gmail.com
[3]Earth and Environmental Systems Institute, The Pennsylvania State University, University Park, PA, USA; sanjibsharma66@gmail.com
[4]The Small Earth Nepal, Kathmandu, Nepal; piyush@smallearth.org.np
[5]Department of Geosciences, University of Rhode Island, Kingston, RI, USA; panthijeeban@gmail.com
[6]Institute of Engineering, Pulchowk Campus, Tribhuvan University, Lalitpur, Nepal; rupesh.baniya480@gmail.com
[7]Bureau of Meteorology, Melbourne, Australia; jayaram.pudashine@bom.gov.au
[8]Universal Engineering and Science College, Nepal; bthapa.ioe@gmail.com
[9]National Research Institute for Earth Science and Disaster Resilience, Tsukuba, Japan; shakti.pc@gmail.com
[10]Department of Hydrology and Meteorology, Ministry of Energy, Water Resources and Irrigation, Kathmandu, Nepal; binodparaj@gmail.com

*Correspondence: binodparaj@gmail.com

**Abstract:**

Extreme rainfall is one of the major causes of natural hazards (for example flood, landslide, and debris flow) in the central Himalayan region, Nepal. The performance of strategies to manage these risks relies on the accuracy of quantitative rainfall estimates. Rain gauges have traditionally been used to measure the amount of rainfall at a given location. The point measurement often misrepresents the basin estimates, because of limited density and high spatial variability of rainfall fields across the Himalayas. The Department of Hydrology and Meteorology (DHM), Nepal has planned to install a network of three weather radars that cover the entire country. So far, the first weather radar has been installed in 2019 in the western region of the country. Two more radars will be added for the planned radar network in the near future covering the central and eastern regions of the country. Here we introduce the first installed weather radar in Nepal. We highlight both the opportunities and challenges with the radar observation in the mountainous regions. Radar rainfall estimates across the Himalayas are critical to issue severe weather warnings; forecast floods and landslides; and inform decision making in a broad range of sectors, including water and energy, construction, transportation, and agriculture.

# 1.0 Introduction

Extreme rainfall-driven hazards such as flash-flood, debris flow and landslide pose major risk to life and property in the Himalayan region, including Nepal. Himalayan river basins are characterised by steep slope, fast runoff processes with short time lag and short response time, and hence the timely and accurate hydrometeorological predictions are quite sensitive to the spatiotemporal variability of rainfall (Habib *et al.*, 2012). In addition, numerical modeling of extreme hydrometeorological events is challenging because of observational constraints (P.C. and Maki, 2017). These hydrometeorological hazards are attributed to either area-specific intense rainfall or large-scale rain-bearing weather systems.

Both climate variability and climate change, expanding urbanization and fragile geological structures are increasing the frequency of damaging natural hazards across the Himalayas (DWIDP, 2014). Nepal has diverse climatic and physiographic regions, ranging from tropical in the Terai Plains to polar on mountain peaks, and passes through temperate hills (Talchabhadel and Karki, 2019). The Northern (hills and *Churia*) regions are more prone to landslides while the Southern region (*Terai*) is more susceptible to floods (Dahal and Hasegawa, 2008). Floods and landslides have caused approximately 8,400 deaths in Nepal from 1983 to 2013, with an average of 269 deaths per year (DWIDP, 2014).

Many of the historical natural disasters are associated with flash-floods, debris flow and landslides triggered by torrential rainfall during the monsoon. Some examples include the 1993 flooding in the central and eastern region, 2008 Koshi embankment breaching, 2012 Seti Kharapani flash flood, 2013 Mahakali flood, 2014 Jure landslide, and 2016 glacial lake outburst flood (GLOF) in Bhotekoshi/Sunkoshi river, 2019 tornado (Qiu, 2016; Cook *et al.*, 2018; DHM *et al.*, 2019; Mallapaty, 2019). A deadly wind storm ripped through the southeastern part of the country on March 31, 2019 killing 28 people and injuring more than 1,100 (DHM *et al.*, 2019; Mallapaty, 2019). Scientists confirmed that it was the country's first ever recorded tornado (Mallapaty, 2019). A recent landslide followed by a debris flow at Kushma municipality, Parbat in central Nepal on June 31, 2020 highlights that the terrain, which never experienced a landslide before, even encountered a disastrous event causing the death of 9 people (NDRRMA, 2020). Increasing frequency and severity of extreme events underscore the pressing need of accurate and reliable quantitative rainfall estimates (Cole and Moore, 2008). For example, reliable spatiotemporal rainfall estimates provide critical inputs to land-surface models for improved monitoring and forecasting of hydrometeorological hazards, including floods and landslides.

Nepal has relied solely on the rain gauge network to measure ground rainfall despite not having a long history of rainfall monitoring. Official monitoring of rainfall started only in 1946 with three rain gauge stations. There have been 23 rainfall stations added in 1947 and 40 more stations in 1950. The hydrometeorological services started systematically only from 1966 and then the number of rain gauges significantly increased till date. The Department of Hydrology and Meteorology (DHM), Nepal is currently operating > 100 automatic and > 400 manual rain gauges distributed across the country. However, the station density in the mountainous region of the country is very sparse. The mountainous region covers approximately 50% of the country's area but only 10% of total stations are located in these remote and rugged areas (Talchabhadel and Karki, 2019).

In general, rainfall variability is monitored using the common rain gauge networks. Two major factors determine the evaluation and design of a rain gauge network: the density of the network and the locations of the rain gauges. A high-density network is needed to obtain accurate information about rainfall distribution. There are several challenges associated with rainfall observations in the Himalayas due to the remote locations involved, the harsh physical environments, the logistical problems and the cost related to maintaining and monitoring the gauges. Meaning, it is likely to miss important structures in the rainfall field in the Himalayan region. Recently, Nepal has started monitoring the lower atmospheric weather condition with Radiosonde technology commonly known as weather balloon in addition to the common rain gauge network (DHM, 2018). Radiosonde monitors the vertical profile of an upper to lower atmosphere and transmits data to a ground based radio station. Radiosonde is useful to track storm-like weather phenomena (Sarkis, 2017).

We know rain gauge informs the amount of rainfall at a particular point and often misrepresents the basin estimates because of sparse density and high spatial variability of rainfall fields (Kidd *et al.*, 2017). Remote sensing technologies are available to address the issue of sparse density of rainfall stations. Satellite-based rainfall estimates (SREs) such as PERSIANN-CCS (~ 4 Km at 1h resolution), and IMERG (~ 10 Km at half h resolution), exhibit the spatial distribution to a larger extent, but demonstrate poor correlation with ground gauges at higher temporal resolution across the Central Himalayas (Shrestha *et al.*, 2012; Alazzy *et al.*, 2017; Zandler *et al.*, 2019). Blending of satellite estimates and gauge rainfall provides relatively better understanding of spatial distribution of rainfall (Thapa *et al.*, 2016). However, the reliability of blending entirely depends on the density of rain gauge stations (Berg *et al.*, 2016; Brocca *et al.*, 2019). Meanwhile, weather radars offer the ability to measure the near real-time rainfall intensity distribution over the wider region (Habib *et al.*, 2012).

This study summarizes the current plan of installation of a network of 3 weather radars, and an overview of a recently installed operational weather radar. We provide an outlook of some representative data of the country's first weather radar installed in the western region of the country. We also highlight the possible implication of these data to different stakeholders, including forecasters, decision-makers and local public. Importantly, we emphasize the direct use of these data in disaster mitigation, preparation and early warning by which many lives could be saved in the first hours which is incomparable with monetary values. The study supports a notion that a dollar spent on disaster preparedness saves multifold costs in disaster response and recovery. The study concludes discussing limitations and challenges in the use of weather radar technology, particularly in mountainous topography of economically developing countries like Nepal, and a way forward.

## 2.0 Weather radar in Nepal

Nepal has planned to install a network of C-band dual polarized weather radars that cover the entire country. So far, the first weather radar has been installed in 2019 at Rata Nangala of Surkhet in western Nepal (**Figure 1**). This radar provides the quantitative rainfall estimates for the western region of the country and has an ability to scan the atmosphere up to 200 km in all directions. The DHM takes care of the operation and maintenance of the installed radar. Two additional radars will be installed in the near future for a planned three-radar network (see **Figure 1**). These additional radars are planned to be installed at Ribdikot of Palpa in the central region and Rametar of Udaipur in the eastern region. The radar technology is useful for short term forecasting (nowcasting) and provides forecasters critical real-time information to enhance warning lead-time when a severe weather situation develops. In addition, DHM uses the radar information to improve the real-time hydrologic monitoring and predictions of different hydrometeorological variables (e.g. rainfall, snowfall) which ultimately provide timely forecast on likely floods, debris flow and landslides in the complex terrain spanning the Himalayan Mountains.

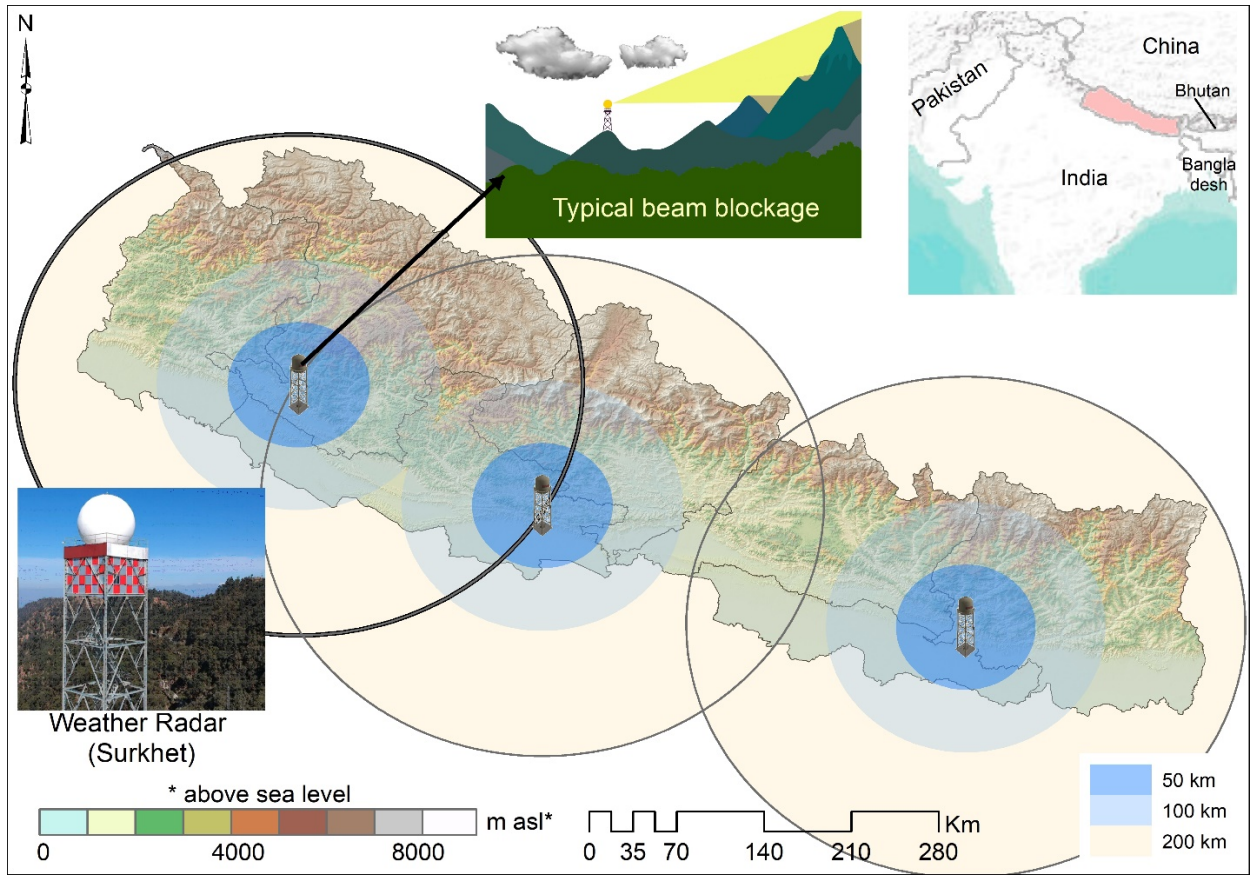

**Figure 1**: Location of three weather radars and their aerial coverages (thick solid line represents operational and thin solid lines represent proposed) across Nepal. The coverage areas are shown in three levels: i) 50 km in dark blue, ii) 100 km in light blue and iii) 200 km in light yellow. Inset at top right shows neighbourhood countries including India, Pakistan, Bangladesh, Bhutan, and China. A typical representation of beak blockage due to the orography, one of the major errors associated with radar observation in mountainous regions, is shown indicating significant blockages around the northern region of Nepal (not in the scale). A location photograph at bottom left shows an operational weather radar (Enterprise Electronics Corporation Defender C350), installed for the first time in Nepal, at Rata Nangla, Surkhet.

**Figures 2a-b** show representative data of the radar reflectivity image from May 05, 2020 for two time slices (15 minutes each) for light- to moderate-rainfall around Surkhet. DHM analyzes radar reflectivity images using standard retrieval algorithms (Krajewski and Smith, 2002) and produces rainfall estimates at each grid point. Currently, the forecasters from DHM perform hybrid (manual + semi-automated) quality control to the radar-rainfall estimates. In addition, DHM analyses different SREs simultaneously in order to strengthen the now-casting. **Figures 2c-d** show

the quantitative rainfall estimates of PERSIANN-CCS at hourly resolution (Hong *et al.*, 2004) at its short time lag (∼1 h). Similarly, **Figures 2e-h** show IMERG Early and Late product at 30-minutes (Huffman *et al.*, 2019) resolution for the aforementioned date. IMERG Early and Late products are available after ~ 4-h and 14-h of observation time, respectively.

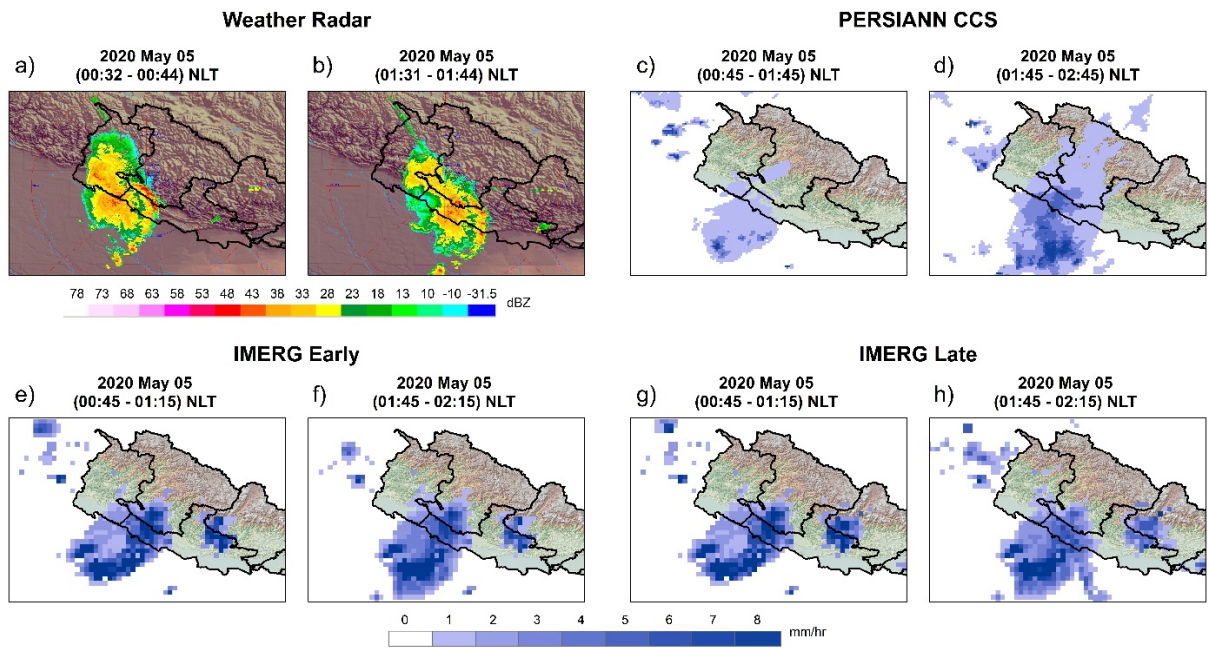

**Figure 2**: Radar reflectivity images for a) May 05, 2020 00:32 AM - 00:44 Nepal Local Time (NLT), and b) May 05, 2020 01:31 AM - 01:44 NLT; and satellite-based rainfall estimates (SREs) derived from PERSIANN-CCS, spatial resolution 4 km, for c) May 05, 2020 00:45 AM - 01:45 NLT, and d) May 05, 2020 01:45 AM - 02:45 NLT; derived from IMERG Early, spatial resolution 10 km, for e) May 05, 2020 00:45 AM - 01:15 NLT, and f) May 05, 2020 01:45 AM - 02:15 NLT; and derived from IMERG-Late, spatial resolution 10 km, for g) May 05, 2020 00:45 AM - 01:15 NLT, and h) May 05, 2020 01:45 AM - 02:15 NLT.

By visually inspecting the weather radar observations (radar reflectivity) and SREs (**Figure 2**), we identify the most salient features. We find a general tendency of different products to capture the rain-bearing systems. However, there are shifts and/or discrepancies in both space and time in SREs. The current finding is entirely based on a single event and does not allow to draw a general conclusion. Indeed, SREs like PERSIANN-CCS and IMERG cannot fully represent the ground rainfall reality on the complex topography of Nepal (Shrestha *et al.*, 2012). However, detailed evaluation of these products using long term continuous data is required to fully understand and characterize the quality of rainfall estimates.

The primary approach for quality control has been to calibrate the radar with rain gauge observations, which are considered as the ground truth. The training and testing of radar products would certainly provide better estimates in the coming days. As of now, the dataset obtained from current weather radar is limited. Long term data is needed to improve the rainfall measurement from radar. Data from rain gauges could be used to assimilate with radar to further improve the accuracy. Also there is a need of thorough and continuous verification of radar data. In the future, the potential of statistical post-processing considering machine/deep learning to remove systematic biases and improve the quality of radar-rainfall estimates could be investigated. At near real-time, the contaminated biases will then be minimized. Consequently, the DHM will be able to numerically quantify the chances of occurrence of flood, landslide and debris flow at specific location and time.

### 3.0 Application of weather radar

The importance of reliable rainfall estimates is well recognized in Nepal (Shrestha *et al.*, 2012). These estimates are useful to improve severe weather detection and warnings, forecast hydrometeorological extremes, and provide real-time hydrologic monitoring (e.g., Krajewski and Smith, 2002; Smith *et al.*, 2002; Krajewski *et al.*, 2010; Reed *et al.*, 2017; Thorndahl *et al.*, 2017), which ultimately are important for mitigation and preparedness of flood, drought, landslides, and debris flow. For instance, the effectiveness of early warning systems for severe hydrometeorological events relies heavily on the adequacy and accuracy of rainfall estimates. However, to take full advantage of high-resolution quantitative rainfall information, it is necessary to improve modeling capacity that allows higher resolution products (Thorndahl *et al.*, 2017). Accurate rainfall estimates in conjunction with high-resolution terrain information, soil moisture data and process-based rainfall-runoff modeling capabilities are required to further enhance the quality of flood forecasts across the central Himalayan region.

Radar rainfall-estimates are useful to inform decision making in a broad range of sectors in Nepal, including water related infrastructure development, energy, construction, transportation, and agriculture. Continuous monitoring of rainfall through the weather radars could improve aviation safety, increase the operational efficiency of the air transport industry, contribute to agriculture advice, help hydropower operations, and plan recreational activities (e.g Smith *et al.*, 2002; EEC, 2020) (see **Figure 3**). For instance, precise and timely rainfall information is important for planning of new hydropower installations, reservoir operation plan and operation of existing ones (e.g., Tapiador *et al.*, 2011). Specifically, for Nepal, early information about the movement and intensity of rainfall help estimate the runoff reaching the hydropower dam/reservoir and adjust its operation to reduce the consequent effects of flooding on people and on the enterprise (CEMIG, 2020). The hydroelectric facilities within the radar domain would benefit from getting more accurate rainfall estimates from radar. For instance, Upper Karnali (900 megawatts), Betan Karnali (688 megawatts), Bheri-4 (300 megawatts), and Bheri-2 (256 megawatts) hydroelectric projects, which have received commercial license for power generation can directly benefit from the radar rainfall estimates at Rata Nangala, Surkhet.

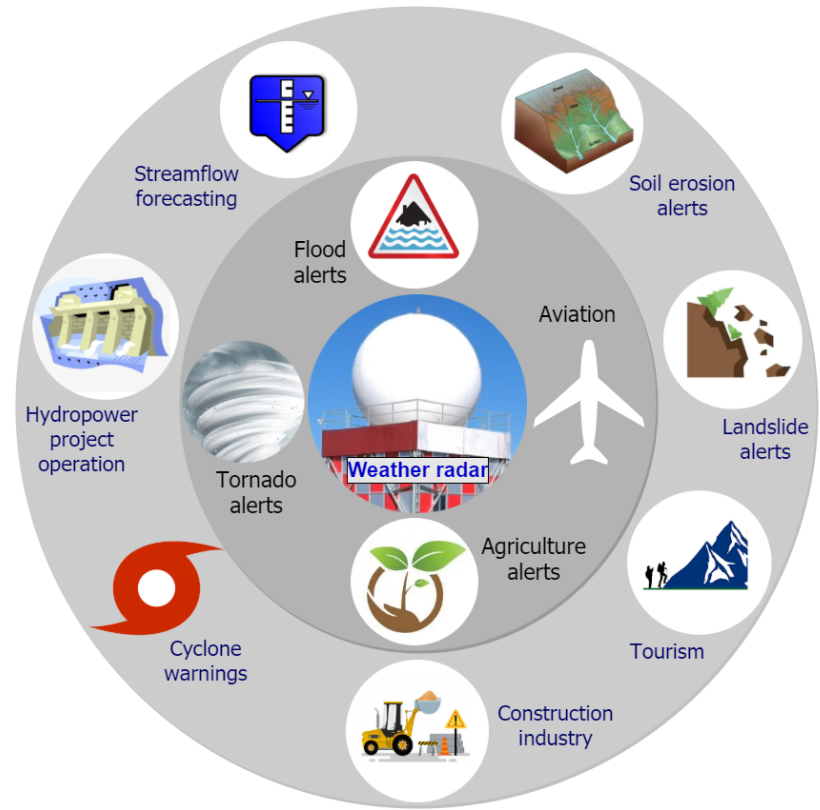

**Figure 3**: Schematic showing various sectors in Nepal that can potentially benefit from radar rainfall estimates. The inner and outer circles show the direct and indirect beneficiaries, respectively of radar rainfall estimates.

### 4.0 Key challenges

Despite recent advances in weather forecasting, accurate and reliable radar rainfall estimation remains a critical issue and challenge. Most notable issues in radar rainfall estimates in the mountainous region are associated with beam blockage followed by signal attenuation, anomalous propagation, bright bands, Z-R relationships, retrieval algorithms and sampling errors (e.g., Wilson and Brandes, 1979; Krajewski and Smith, 2002; Krajewski *et al.*, 2010; P.C. and Maki, 2017; Reed *et al.*, 2017; Ryzhkov and Zrnic 1995; Carey et al., 2000; Park et al., 2005; Kim et al., 2010; Maki et al. 2012). For instance, one would expect to experience a prominent challenge in handling the uncertainties from beam blockage given the unique basin topography mostly characterized by a complex and rugged mountains, pocket valleys, and extraordinary elevation range (P.C. et al., 2013). There are a number of additional challenges with radar rainfall observation in Nepal, including in-situ instrument failure due to power outage; lack of technical expertise for regular maintenance and data collection, data storage, processing and dissemination; and financial capability to install and operate weather radars. On top of it, there is a big challenge of transboundary data sharing in the central Himalayan region (Akanda, 2012; Singh and Thadani, 2015).

### 5.0 Way forward

In this study, we present the first ever installed weather radar in Nepal. The study highlights both the opportunities and challenges with the radar system in the central Himalayan region. Radar rainfall estimates across mountainous topography are critical to issue severe weather warnings; forecast floods and landslides; and inform decision making in a broad range of sectors, including energy, construction, transportation, and agriculture. There is a huge opportunity to enhance the hydrometeorological forecasting capabilities in Nepal through integrated observations from currently available systems. The combination of rainfall information from satellite, radar and weather stations can enhance the quality of quantitative rainfall estimates. These estimates can be used to enhance the physical realisms of real-time watershed monitoring, improve hazard predictions, communicate hazard information, and enhance emergency planning and response. There are several options available for estimating the spatial and temporal distribution of rainfall including the most common types of weather radar, which employ C-, S-, or X-band wavelengths.

Natural hazards such as floods pose major risks to life and property in Nepal. These risks are expected to increase in the future with demographic, environmental and urbanization changes. High-spatiotemporal resolution radar rainfall estimates at urban areas such as that from X-band polarimetry radar (e.g., Krajewski and Smith, 2002; Berne and Krajewski, 2013) can complement flood preparedness efforts in Nepal. In addition, effective implementation of natural hazards risk management strategies requires coordinated efforts among the research, operational and management communities towards developing a cyberinfrastructure that integrate observations, high-performance computing, early warning systems, transparent information management and advanced communication services in a networked environment. Hence, mutual collaboration or sharing knowledge in weather radar observation could be helpful to enhance the quantitative precipitation estimation over the central Himalayan region.

**Funding:** This research received no external funding.

**Acknowledgments:** The authors greatly acknowledge the Department of Hydrology and Meteorology (DHM), Nepal for providing the valuable information regarding the weather radar. The authors are grateful to the anonymous reviewers for their reviews and constructive comments.

**Conflicts of Interest:** The authors declare no conflict of interest.

**References:**

Akanda AS. 2012. South Asia's water conundrum: hydroclimatic and geopolitical asymmetry, and brewing conflicts in the Eastern Himalayas. *International Journal of River Basin Management*, 10(4): 307–315. https://doi.org/10.1080/15715124.2012.727824.

Alazzy AA, Lü H, Chen R, Ali AB, Zhu Y, Su J. 2017. Evaluation of Satellite Precipitation Products and Their Potential Influence on Hydrological Modeling over the Ganzi River Basin of the Tibetan Plateau. *Advances in Meteorology*, 2017. https://doi.org/10.1155/2017/3695285.

Berg P, Norin L, Olsson J. 2016. Creation of a high resolution precipitation data set by merging gridded gauge data and radar observations for Sweden. *Journal of Hydrology*. Elsevier B.V., 541: 6–13. https://doi.org/10.1016/j.jhydrol.2015.11.031.

Berne A, Krajewski WF. 2013. Radar for hydrology: Unfulfilled promise or unrecognized potential? *Advances in Water Resources*, 51: 357–366. https://doi.org/10.1016/j.advwatres.2012.05.005.

Brocca L, Filippucci P, Hahn S, Ciabatta L, Massari C, Camici S, Schüller L, Bojkov B, Wagner W. 2019. SM2RAIN–ASCAT (2007–2018): global daily satellite rainfall data from ASCAT soil moisture observations. *Earth System Science Data*, 11(4): 1583–1601. https://doi.org/10.5194/essd-11-1583-2019.

Carey LD, Rutledge SA, Ahijevych DA, and Keenan TD. 2000. Correcting propagation effects in C-band polarimetric radar observations of tropical convection using differential propagation phase. Journal of Applied Meteorology 39: 1405-1433.

CEMIG. 2020. *Meteorological radar*. Available at: https://www.cemig.com.br/en-us/Company_and_Future/Sustainability/water_resources/Pages/meteorological_radar.aspx.

Cole SJ, Moore RJ. 2008. Hydrological modelling using raingauge- and radar-based estimators of areal rainfall. *Journal of Hydrology*, 358(3–4): 159–181. https://doi.org/10.1016/j.jhydrol.2008.05.025.

Cook KL, Andermann C, Gimbert F, Adhikari BR, Hovius N. 2018. Glacial lake outburst floods as drivers of fluvial erosion in the Himalaya. *Science*, 362(6410): 53–57. https://doi.org/10.1126/science.aat4981.

Dahal RK, Hasegawa S. 2008. Representative rainfall thresholds for landslides in the Nepal Himalaya. *Geomorphology*, 100(3–4): 429–443. https://doi.org/10.1016/j.geomorph.2008.01.014.

DHM 2018. *Press release on radiosonde*. Kathmandu, Nepal; can be assessed at https://www.google.com/url?q=https://www.dhm.gov.np/uploads/getdown/1454071866DHM.pdf&sa=D&ust=1595136158812000&usg=AFQjCNHicLKVI9y9wU4IEqYw0Uu25dWQiA

DHM, SEN, ICIMOD. 2019. *Report on Bara-Parsa Tornado*. Kathmandu, Nepal; can be assessed at :http://www.smallearth.org.np/wp-content/uploads/2019/04/Report-on-Bara-Parsa-Tornado.pdf

DWIDP. 2014. *Disaster Review 2013*. Kathmandu, Nepal; can be assessed at: https://reliefweb.int/sites/reliefweb.int/files/resources/DWIDP_Review_2013_Book_All_Pages_Final.pdf

EEC. 2020. Enterprise Electronics Corporation (EEC) Inks Deal to Expand Weather Radar Network in Nepal. Kathmandu, Nepal; can be assessed at: http://www.eecradar.com/pdf/Nepal-Press-Release.pdf

Habib E, Haile AT, Tian Y, Joyce RJ. 2012. Evaluation of the high-resolution CMORPH satellite rainfall product using dense rain gauge observations and radar-based estimates. *Journal of Hydrometeorology*, 13(6): 1784–1798. https://doi.org/10.1175/JHM-D-12-017.1.

Hong Y, Hsu K-L, Sorooshian S, Gao X. 2004. Precipitation Estimation from Remotely Sensed Imagery Using an Artificial Neural Network Cloud Classification System. *Journal of Applied Meteorology*, 43(12): 1834–1853. https://doi.org/10.1175/JAM2173.1.

Huffman GJ, Bolvin DT, Braithwaite D, Hsu K-L, Joyce R, Kidd C, Nelkin EJ, Sorooshian S, Tan J, Xie P. 2019. *Algorithm Theoretical Basis Document (ATBD) Version 06 NASA Global Precipitation Measurement (GPM) Integrated Multi-satellitE Retrievals for GPM (IMERG)*. .

Kidd C, Becker A, Huffman GJ, Muller CL, Joe P, Skofronick-Jackson G, Kirschbaum DB. 2017. So, How Much of the Earth's Surface Is Covered by Rain Gauges? *Bulletin of the American Meteorological Society*, 98(1): 69–78. https://doi.org/10.1175/BAMS-D-14-00283.1.

Kim DS, Maki M, Lee DI. 2010. Retrieval of three-dimensional raindrop size distribution using X-band polarimetric radar data. *Journal of Atmospheric and Oceanic Technology* 27: 1265-1285.

Krajewski WF, Smith JA. 2002. Radar hydrology: rainfall estimation. *Advances in Water Resources*, 25(8–12): 1387–1394. https://doi.org/10.1016/S0309-1708(02)00062-3.

Krajewski WF, Villarini G, Smith JA. 2010. RADAR-Rainfall Uncertainties. *Bulletin of the American Meteorological Society*, 91(1): 87–94. https://doi.org/10.1175/2009BAMS2747.1.

Kubota T, Ushio T, Shige S, Kida S, Kachi M, Okamoto K. 2009. Validation of high-resolution satellite-based rainfall estimates around Japan using gauge-calibrated ground-radar dataset. *Journal of Meteorological Society Japan* 87: 203-222.

Maki M, Maesaka T, Kato A, Kim DS, Iwanami K. 2012. Composite rainfall map with X-band polarimetric radar network and C-band conventional radar. *Indian J Radio Space Phys* 41: 461-470.

Mallapaty S. 2019. Nepali scientists record country's first tornado. *Nature*.

NDRRMA. 2020. *Report on Parbat Durlung Landslide*. Kathmandu, Nepal; can be assessed at http://drrportal.gov.np/uploads/document/1633.pdf (in Nepali Language)

Park SG, Maki M, Iwanami K, Bringi VN, Chandrasekar V. 2005. Correction of radar reflectivity and differential reflectivity for rain attenuation at X band. Part II: Evaluation and application. *Journal of Atmospheric and Oceanic Technology* 22: 1633-1655

P.C., S, Maki M, Shimizu S, Maesaka T, Kim D, Lee D, Iida H. 2013. Correction of Reflectivity in the Presence of Partial Beam Blockage over a Mountainous Region Using X-Band Dual Polarization Radar. *J. Hydrometeor.*, **14**, 744–764, https://doi.org/10.1175/JHM-D-12-077.1.

P.C. S, Misumi R, Nakatani T, Iwanam K, Maki M, Maesaka T, Hirano K. 2016. Accuracy of Quantitative Precipitation Estimation Using Operational Weather Radars: A Case Study of Heavy Rainfall on 9–10 September 2015 in the East Kanto Region, Japan. *J. Disaster Res.*, 11(5):1003-1016. https://doi.org/10.20965/jdr.2016.p1003.

P.C. S, Maki M. 2017. Challenges in Estimating Quantitative Precipitation Estimation (QPE) Using Weather Radar Observation Over the Mountainous Country of Nepal. *Hydro Nepal: Journal of Water, Energy and Environment*, 21(21): 50–59. https://doi.org/10.3126/hn.v21i0.17822.

P.C. Shakti, Nakatani T, Misumi R. 2018. Analysis of Flood Inundation in Ungauged Mountainous River Basins: A Case Study of an Extreme Rain Event on 5–6 July 2017 in Northern Kyushu, Japan. *J. Disaster Res.*, 13(5): 860-872. https://doi.org/10.20965/jdr.2018.p0860.

P.C., S.; Nakatani, T.; Misumi, R. 2019. The Role of the Spatial Distribution of Radar Rainfall on Hydrological Modeling for an Urbanized River Basin in Japan. *Water*, *11*, 1703. https://doi.org/10.3390/w11081703

P.C., S.; Kamimera, H.; Misumi, R. 2020. Inundation Analysis of the Oda River Basin in Japan during the Flood Event of 6–7 July 2018 Utilizing Local and Global Hydrographic Data. *Water*, *12*, 1005. https://doi.org/10.3390/w12041005

Qiu J. 2016. Killer landslides: The lasting legacy of Nepal's quake. *Nature*, 532(7600): 428–431. https://doi.org/10.1038/532428a.

Reed JL, Lanterman AD, Trostel JM. 2017. Weather radar: Operation and phenomenology. *IEEE Aerospace and Electronic Systems Magazine*, 32(7): 46–62. https://doi.org/10.1109/MAES.2017.150178.

Ryzhkov A, Zrnic DS. 1995. Precipitation and attenuation measurements at a 10-cm wavelength. *Journal of Applied Meteorology* 34: 2121-2134.

Shrestha D, Singh P, Nakamura K. 2012. Spatiotemporal variation of rainfall over the central Himalayan region revealed by TRMM Precipitation Radar. *Journal of Geophysical Research: Atmospheres*, 117(D22). https://doi.org/10.1029/2012JD018140.

Sarkis S. 2017. Upper Air Observations: How Weather Balloons Improve Forecasts. Available at: https://www.aoml.noaa.gov/keynotes/keynotes_0917_weatherballoons.html

Singh SP, Thadani R. 2015. Complexities and Controversies in Himalayan Research: A Call for Collaboration and Rigor for Better Data. *Mountain Research and Development*, 35(4): 401–409. https://doi.org/10.1659/MRD-JOURNAL-D-15-00045.

Smith PL, Atlas D, Bluestein HB, Chandrasekar V, Kalnay E, Keeler RJ, McCarthy J, Rutledge SA, Seliga TA, Serafin RJ, Wilson FW, Turekian VC. 2002. *Weather Radar Technology Beyond NEXRAD*. National Academies Press: Washington, D.C.

Talchabhadel R, Karki R. 2019. Assessing climate boundary shifting under climate change scenarios across Nepal. *Environmental Monitoring and Assessment*. Environmental Monitoring and Assessment, 191(8): 520. https://doi.org/10.1007/s10661-019-7644-4.

Tapiador FJ, Hou AY, de Castro M, Checa R, Cuartero F, Barros AP. 2011. Precipitation estimates for hydroelectricity. *Energy & Environmental Science*, 4(11): 4435. https://doi.org/10.1039/c1ee01745d.

Thapa BR, Ishidaira H, Bui TH, Shakya NM. 2016. Evaluation of water resources in mountainous region of Kathmandu Valley using high resolution satellite precipitation product. *Journal of Japan Society of Civil Engineers, Ser. G (Environmental Research)*, 72(5): I_27-I_33. https://doi.org/10.2208/jscejer.72.I_27.

Thorndahl S, Einfalt T, Willems P, Nielsen JE, ten Veldhuis M-C, Arnbjerg-Nielsen K, Rasmussen MR, Molnar P. 2017. Weather radar rainfall data in urban hydrology. *Hydrology and Earth System Sciences*, 21(3): 1359–1380. https://doi.org/10.5194/hess-21-1359-2017.

Wilson JW, Brandes EA. 1979. Radar Measurement of Rainfall—A Summary. *Bulletin of the American Meteorological Society*, 60(9): 1048–1058. https://doi.org/10.1175/1520-0477(1979)060<1048:RMORS>2.0.CO;2.

Zandler H, Haag I, Samimi C. 2019. Evaluation needs and temporal performance differences of gridded precipitation products in peripheral mountain regions. *Scientific Reports*. Springer US, 9(1): 1–15. https://doi.org/10.1038/s41598-019-51666-z.